\documentclass[12pt,journal,final,onecolumn]{IEEEtran}

\usepackage[dvips]{graphicx}
\usepackage{epsfig}
\usepackage[cmex10]{amsmath}
\usepackage{amssymb}
\usepackage{amsthm}
\usepackage{amsfonts}
\usepackage{bm}
\usepackage{cite}
\usepackage[svgnames]{xcolor} 
\usepackage{pstricks,pst-node,pst-plot,pstricks-add}
\usepackage[tight,footnotesize]{subfigure}
\usepackage{algorithm}
\usepackage{algpseudocode}

\graphicspath{{figs/}}

\input{niesen.def}

\begin{document}

\title{Online Coded Caching}

\author{Ramtin Pedarsani, Mohammad Ali Maddah-Ali, and Urs Niesen%
    \thanks{R. Pedarsani is with UC Berkeley, M. A. Maddah-Ali and U.
        Niesen are with Bell Labs, Alcatel-Lucent. Emails:
        ramtin@eecs.berkeley.edu,
        mohammadali.maddah-ali@alcatel-lucent.com,
        urs.niesen@alcatel-lucent.com. This work was done when R. Pedarsani
        was a summer intern at Bell Labs, Alcatel-Lucent.}%
    \thanks{The work of U. Niesen was supported in part by AFOSR under grant FA9550-09-1-0317.}%
}

\maketitle

\begin{abstract} 
    We consider a basic content distribution scenario consisting of a
    single origin server connected through a shared bottleneck link to a
    number of users each equipped with a cache of finite memory. The
    users issue a sequence of content requests from a set of popular
    files,  and the goal is to operate the caches as well as the server
    such that these requests are satisfied with the minimum number of
    bits sent over the shared link.  Assuming a basic Markov model for
    renewing the set of popular files, we characterize approximately the
    optimal long-term average rate of the shared link.  We further prove
    that the optimal online scheme has approximately the same
    performance as the optimal offline scheme, in which the cache
    contents can be updated based on the entire set of popular files
    before each new request. To support these theoretical results, we
    propose an online coded caching scheme termed \emph{coded
    least-recently sent} (LRS) and simulate it for a demand time series
    derived from the dataset made available by Netflix for the Netflix
    Prize. For this time series, we show that the proposed coded LRS
    algorithm significantly outperforms the popular least-recently used
    (LRU) caching algorithm.
\end{abstract}

\section{Introduction}
\label{sec:intro}

The demand for video streaming services such as those offered by
Netflix, YouTube, Amazon, and others, is growing rapidly. This places a
significant burden on networks. One way to mitigate this burden is to
place memories into the network that can be used to cache files that
users may request. In this paper, we investigate how to optimally use
these caches. In particular, we are interested in \emph{online}
algorithms for this problem, in which the operations of the cache have
to be performed on the fly and without knowledge of future requests.

The online caching problem (also known as the paging problem in the
context of virtual memory systems) has a long history, dating back to
the work by Belady in 1966~\cite{belady66}. This problem has been
investigated both for systems with a single cache~\cite{belady66,
sleator85, fiat91, mcgeoch91, borodin91, karlin92, young94, cao97,
chrobak98, young98, torng98, koutsoupias00, angelopoulos07} as well as
for systems with multiple distributed caches \cite{awerbuch96, heide00,
li06}. One solution to the caching problem that is popular in practice
and for which strong optimality guarantees can be
proved~\cite{sleator85, borodin91, chrobak98, torng98, koutsoupias00,
angelopoulos07} is the \emph{least-recently used} (LRU) eviction policy.
In LRU, each cache is continuously updated to hold the most recently
requested files, allowing it to exploit the temporal locality of content
requests. 

The figure of merit adopted by the papers mentioned so far is the
cache-miss rate (or page-fault rate in the context of the paging
problem), sometimes weighted by the file size. This cache-miss rate is
used as a proxy for the network load. For systems with a \emph{single}
cache, the weighted cache-miss rate and the network load are indeed
proportional to each other, and hence minimizing the former also
minimizes the latter. However, this proportionality no longer holds for
systems with \emph{multiple} caches. For such systems with multiple
caches, a fundamentally different so-called \emph{coded caching}
approach is required. This coded caching approach has been recently
introduced in~\cite{maddah-ali12a, maddah-ali13, niesen13} for the
\emph{offline} caching problem.

\begin{figure}[htbp]
    \centering
    \includegraphics{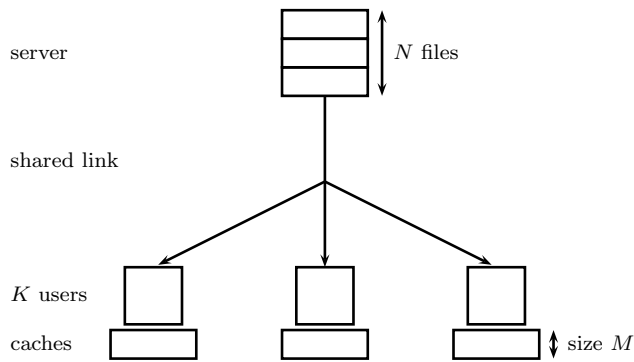}
    \caption{Caching system considered in this paper. A
    server containing $N$ files of size $F$ bits each is connected
    through a shared link to $K$ users each with a cache of size $MF$
    bits. In the figure, $N=K=3$ and $M=1$.} 
    \label{fig:setting}
\end{figure}

In this paper, we investigate \emph{online} coded caching, focusing on a
basic content distribution scenario consisting of a single origin server
connected through a shared (bottleneck) link to a number of users each
equipped with a cache of finite memory (see Fig.~\ref{fig:setting}). The
users issue a sequence of content requests from a set of popular files,
and the goal is to operate the caches as well as the server such as to
satisfy these requests with the minimum number of bits sent over the
shared link. We consider the case where the set of popular files evolve
according to a Markov model and users select their demand uniformly from
this set. 

We approximately characterize the optimal long-term average rate of the
shared link for this setting. We show further that the optimal online
scheme performs approximately the same as the optimal offline scheme.
This is perhaps surprising, since in the offline scheme caches are
allowed to be updated in an offline fashion each time the set of popular
files changes, whereas in the online scheme caches are updated in an
online fashion based solely on the limited observations they have
through the sequence of requests.\footnote{This definition of an
offline caching scheme differs from the definition adopted by other
papers in the caching literature such as~\cite{sleator85}. In those
papers, an offline scheme is one that has noncausal access to the 
entire sequence of demands.}

To evaluate the gain of coded caching in practical scenarios, we propose
an online coded caching scheme termed \emph{coded least-recently sent}
(LRS) and simulate it on a demand time series derived from the dataset
made available by Netflix for the Netflix Prize. For this time series,
we show that the proposed coded LRS algorithm significantly outperforms
the baseline LRU algorithm.  

The remainder of this paper is organized as follows.
Section~\ref{sec:background} provides background information on coded
caching. Section~\ref{sec:problem} formally introduces the problem
setting. Section~\ref{sec:main} contains the main results.  The proof of
these results are provided in Section \ref{sec:proof1}.

\section{Background on Coded Caching}
\label{sec:background}

Coded caching, recently introduced in \cite{maddah-ali12a, maddah-ali13,
niesen13}, is a novel approach to the distributed caching problem. It
can achieve a significant reduction in network load by creating and
exploiting coded multicasting opportunities between users with
different demands. We make essential use of the offline coded caching
scheme from~\cite{maddah-ali13} in the present paper. Therefore, we now
briefly overview that algorithm and illustrate it with an example. 

The setting in~\cite{maddah-ali13} is the offline version of the one
depicted in Fig.~\ref{fig:setting} in Section~\ref{sec:intro}. In
particular, a single origin server is connected to $K$ users through a
shared link. There is a fixed set of $N\geq K$ files of length $F$ bits,
and each user has a memory of size $MF$ bits with $M \leq N$. The cache
memories are prefetched in an \emph{offline} fashion during a placement
phase (during a period of low network load, say the early morning) so as
to minimize the peak load $R(M,N,K)F$ over the shared link during a
later delivery phase (say in the evening) during which each user
requests a single file. We refer to the normalized peak load $R(M,N,K)$
as the peak rate.

The offline coded caching scheme proposed in~\cite{maddah-ali13}
achieves a peak rate of 
\begin{equation}
    \label{eq:dec}
    R(M, N, K) \defeq
    K\cdot(1-M/N)\cdot\frac{N}{KM}\bigl(1-(1-M/N)^K\bigr),
\end{equation}
which is shown to be within a constant factor of the optimal rate.

In the placement phase of the algorithm in~\cite{maddah-ali13}, each
user caches a random subset of $MF/N$ bits of each of the $N$ files. In
the delivery phase, the server sends an appropriate linear combination
of those bits over the shared link to enable all users to recover the
requested files, as is illustrated in the following example.

\begin{example}[\emph{Decentralized Caching Scheme~\cite{maddah-ali13}}]
    \label{eg:dec}
    Consider the caching problem with $N=2$ files say $A, B$ and $K=2$
    users, each with a cache of size $MF$. In the placement phase, each
    user caches $MF/2$ bits of each file independently at random,
    satisfying the memory constraint. We partition
    \begin{equation*}
        A = (A_{\emptyset},A_1,A_2,A_{1,2}),
    \end{equation*}
    where $A_\mc{S}$ denotes the bits of file $A$ that are stored at the
    users in the set $\mc{S}$, and similarly for $B$. For $F$ large
    enough, the size of the subfile $A_\mc{S}$ tends to
    $(M/2)^{\card{\mc{S}}}(1-M/2)^{2-\card{\mc{S}}}F$ bits by the law of
    large numbers. 
    
    In the delivery phase, suppose that users one and two request files
    $A$ and $B$, respectively. User one already has access to the file
    parts $A_{1}$ and $A_{1,2}$ of its requested file, and needs
    $A_{\emptyset}$ and $A_2$, which are not cached its memory.
    Similarly user two already has access to $B_{2}$ and $B_{1,2}$ of
    its requested file and needs $B_{\emptyset}$ and $B_2$. The server
    can then satisfy these user requests by sending $A_{\emptyset}$,
    $B_{\emptyset}$, and $A_2 \oplus B_1$ over the shared link, where
    $\oplus$ denotes the XOR operation applied element-wise to to $A_2$
    and $B_1$ treated as vectors of bits.

    Observe that user one has $B_1$ stored in its cache. From this and
    the output $A_2 \oplus B_1$ of the shared link, user one can recover
    the desired file part $A_2$. Similarly, user two has $A_2$ stored in
    its cache and can use this to recover the desired file part $B_1$
    from the output $A_2 \oplus B_1$ of the shared link.  Thus, using
    the contents of their caches and the outputs of the shared link,
    both user can recover all the required subfiles. 
    
    The rate over the shared link is
    \begin{equation*}
        (M/2)(1-M/2)+2(1-M/2)^2 = R(M,2,2),
    \end{equation*}
    where $R(M,N,K)$ is defined in~\eqref{eq:dec}. While here the
    delivery phase is explained for a specific set of user requests, one
    can verify that this rate is achievable for all other possible
    requests as well. 
\end{example}

The main gain from using this scheme derives from the coded multicasting
opportunities between users with different demands. These coded
multicasting opportunities are created in the placement phase and are
exploited in the delivery phase. As the size $M$ of the cache memories
increases, this coded multicasting gain increases as well. This gain,
called the \emph{global} gain in~\cite{maddah-ali12a, maddah-ali13}, is
captured by the factor 
\begin{equation*}
    \frac{N}{KM}\bigl(1-(1-M/N)^K\bigr)
\end{equation*}
in $R(M,N,K)$. There is a second, \emph{local}, gain deriving from
having part of the requested file available at a users local cache.
This local gain is captured by the factor
\begin{equation*}
    (1-M/N)
\end{equation*}
in $R(M,N,K)$.  As is shown in~\cite{maddah-ali12a, maddah-ali13,
niesen13}, this local gain is usually less significant than the global
coded multicasting gain.

\section{Problem Setting}
\label{sec:problem}

We consider a content distribution system with a server connected
through a shared, error-free link to $K$ users as illustrated in
Fig.~\ref{fig:setting} in Section~\ref{sec:intro}. At time $t\in\N$,
each user $k$ requests a file $d_t(k)$ from a time-varying set
$\mc{N}_t$ of popular files with cardinality $N\geq K$. The $K$ user's
requests, collectively denoted by the vector $d_t$, are chosen uniformly
at random without replacement from $\mc{N}_t$.  Each file has size $F$
bits, and each user is equipped with a cache memory of size $MF$ bits. 

The content distribution system operates as follows. At the beginning of
each time slot $t$, the users reveal their requests $d_t$ to the server.
The server, having access to the database of all the files in the
system, responds by transmitting a message of size $R_tF$ bits over the
shared link. Using their cache contents and the message received over
the shared link, each user $k$ aims to reconstruct its requested file
$d_t(k)$. 

The goal is to design the actions of the users and the server such as to
minimize the long-term average rate $\bar{R}$ of the system, i.e., 
\begin{equation}
    \label{eq:ar}
    \bar{R} \defeq \limsup_{T\to\infty} \frac{1}{T}\sum_{t=1}^T\E(R_t),
\end{equation}
while satisfying the memory and reconstruction constraints.  Observe
that the rate $\bar{R}$ is the long-term average load $\bar{R}F$ over
the shared link normalized by the file size $F$. In order to obtain a
rate $\bar{R}$ independent of the file size and to simplify the
analysis, we allow the file size $F$ to be as large as needed. 

In this paper, we are interested in \emph{online} caching schemes, which
place additional restrictions on the actions of the server and the
caches. In such online schemes, the cache content at user $k$ at the
beginning of time slot $t$ is a function of the cache content at the
same user at the previous time $t-1$, the output of the shared link at
time $t-1$, and the requests $d_1, d_2, \dots, d_{t-1}$ up until time
$t-1$.  In particular, the cache content may \emph{not} be a function of
the outputs of the shared link at times prior to $t-1$. Furthermore, for
an online scheme, the message sent by the server over the shared link at
time $t$ is a function of only the demands $d_{t}$ and the cache
contents of the users at that time. We denote by $\bar{R}^\star$ the
long-term average rate over the shared link of the optimal online
caching scheme. 

\begin{example}[\emph{LRU}]
    \label{eg:lru}
    A popular online caching scheme is \emph{least-recently used} (LRU).
    In this scheme, each user caches $M$ entire files. When a user
    requests a file that is already contained in its cache, that request
    can be served out of the cache without any communication from the
    server. When a user requests a file that is not contained in the
    cache, the server sends the entire file over the link. The user then
    evicts the least-recently requested (or used) file from its cache
    and replaces it with the newly requested one.

    Observe that this is a valid online caching strategy. Indeed, the
    content of a user's cache at the beginning of time slot $t$ is a
    function of the cache content at time $t-1$, the output of the
    shared link at time $t-1$, and the past requests (in order to
    determine which file was least-recently used). Moreover, the message
    sent by the server at time $t$ is only a function of the demands
    $d_t$ and the cache contents of the users  at that time (in order to
    decide if a file needs to be transmitted at all). We will adopt LRU
    as the baseline scheme throughout this paper.
\end{example}

We next provide a formal description of the dynamics of the set of
popular files $\mc{N}_t$. The initial set $\mc{N}_1$ consists of $N$
distinct files. The set $\mc{N}_{t}$ at time $t$ evolves from the set
$\mc{N}_{t-1}$ at time $t-1$ using an arrival/departure process.  With
probability $1-p$, there is no new arrival and $\mc{N}_{t} =
\mc{N}_{t-1}$. With probability $p$, there is a new arrival, and the set
$\mc{N}_{t}$ is constructed by choosing a file uniformly at random from
$\mc{N}_{t-1}$ and replacing it with a new, so far unseen, file.  Note
that this guarantees that $\card{\mc{N}_t} = N$ for all $t$. 

\begin{example}[\emph{Popular-File Dynamics}]
    \label{eg:popular}
    Consider a system with $N=2$ popular files. A possible evolution of
    the set $\mc{N}_t$ of popular files is as follows.
    \begin{description}
        \item[$t=1$:] The initial set of popular files is $\mc{N}_1 = \{B, C\}$. 
        \item[$t=2$:] There is an arrival. The file $C$ is randomly
            chosen and replaced with the new file $D$, so that
            $\mc{N}_2 = \{B, D\}$.
        \item[$t=3$:] There is no arrival, and $\mc{N}_3 = \mc{N}_2 =
            \{B, D\}$. 
    \end{description}
\end{example}

\section{Main Results}
\label{sec:main}

We start by introducing a new, online, coded caching algorithm in
Section~\ref{sec:main_algo}. A simplified variant of this algorithm is
shown in Section~\ref{sec:main_theory} to have performance close to the
optimal online caching scheme. Section~\ref{sec:main_sim} provides
simulation results comparing the proposed coded caching algorithm to the
baseline LRU scheme for an empirical demand time series derived from the
Netflix Prize database.

\subsection{An Online, Coded Caching Algorithm}
\label{sec:main_algo}

In this section, we propose an online version of the caching algorithm
in \cite{maddah-ali13}, which we term \emph{coded least-recently sent}
(LRS).  The coded LRS algorithm is presented formally in the listing
Algorithm~\ref{alg:1}. The statement of the algorithm uses the shorthand
$[K]$ for $\{1,2,\dots,K\}$. 

\begin{algorithm}[htbp!]
    \caption{The coded LRS caching algorithm for time $t$.}
    \label{alg:1}
    \begin{algorithmic}[1]
        \Procedure{Delivery}{}
        \For{$s=K, K-1, \ldots, 1$} \label{alg:1_sloop}
        \For{$\mc{S}\subset[K]: \card{\mc{S}}=s$} \label{alg:1_Sloop}
        \State Server sends \(\oplus_{k\in\mc{S}} V_{\mc{S}\setminus\{k\}}(k)\) \label{alg:send}
        \EndFor 
        \EndFor 
        \EndProcedure
        \Procedure{Cache Update}{}
        \For{$k, k' \in[K]$}
        \If{$d_t(k')$ is not partially cached at user $k$} 
        \State User $k$ replaces the least recently sent file in its
        cache with a random subset of $\tfrac{MF}{N'}$ bits of file $d_t(k')$  \label{alg:1_cache}
        \EndIf
        \EndFor
        \EndProcedure
    \end{algorithmic}
\end{algorithm} 

Algorithm~\ref{alg:1} consists of a delivery procedure and a cache
update procedure. We now explain those two procedures in detail. The
delivery procedure is formally similar to the delivery procedure of the
decentralized caching algorithm in \cite{maddah-ali13}. $V_{\mc{S}}(k)$
denotes the bits of the file $d_t(k)$ requested by user $k$ cached
exclusively at users in $\mc{S}$. In other words, a bit of file $d_t(k)$
is in $V_{\mc{S}}(k)$ if it is present in the cache of every user in
$\mc{S}$ and if it is absent from the cache of every user outside
$\mc{S}$. The XOR operation $\oplus$ in Line~\ref{alg:send} is to be
understood as being applied element-wise to $V_{\mc{S}}(k)$ treated as a
vector of bits. If those vectors do not have the same length, they are
assumed to be zero padded for the purposes of the XOR. We thus see that
the delivery procedure of the coded LRS algorithm consists of sending
one linear file combination for each subset $\mc{S}$ of users. It is
worth pointing out that, whenever a requested file is not cached at any
user, then $V_{\emptyset}(k)$ is equal to the entire requested file, and
hence when $\mc{S} = \{k\}$ in Line~\ref{alg:1_Sloop} the delivery
procedure sends in this case the entire requested file uncoded over the
shared link.

Consider next the cache update procedure. In each time slot $t$, the
users maintain a list of 
\begin{equation*}
    N' \defeq \alpha N
\end{equation*}
partially cached files for some $\alpha \geq 1$. The parameter $\alpha$
can be chosen to optimize the caching performance; a simple and
reasonable choice is $\alpha=1.4$. At time $t$, after the delivery
procedure is executed, the caches are updated as follows. If a requested
file $d_t(k')$ of \emph{any} user $k'$ is not currently partially
cached, \emph{all} users evict the least-recently used file and replace
it with $MF/N'$ randomly chosen bits from file $d_t(k')$. This is
feasible since the uncached file $d_t(k')$ was sent uncoded over the
shared link during the delivery procedure. Note that this update
procedure guarantees that the number of partially cached files remains
$N'$, and that the same files (but not necessarily the same bits) are
partially cached at each of the $K$ users.

We illustrate the proposed coded LRS algorithm with an example. This
example also illustrates that the rate of the proposed scheme can be
related to the rate $R(M,N,K)$ defined in~\eqref{eq:dec} of the
decentralized caching algorithm from~\cite{maddah-ali13}.

\begin{example}[\emph{Coded LRS}] 
    We consider again a system with $N=2$ popular files and assume the
    same popular-file dynamics as in Example~\ref{eg:popular} in
    Section~\ref{sec:problem}. Assume there are $K=2$ users with a cache
    memory of $M=1$. Let $\alpha=3/2$ so that each user caches a
    fraction $1/3$ of $N' = \alpha N = 3$ files. We assume that
    initially each user partially caches the files $\{A, B, C\}$.
    \begin{description}
        \item[$t=1$:] The set of popular files is $\mc{N}_1 = \{B, C\}$.
            Assume the users request $d_1 = (B,C)$. Both of the
            requested files are partially cached at the users. In the
            delivery procedure, the server sends $B_\emptyset$,
            $C_\emptyset$, and $B_2\oplus C_1$. For $F$ large enough, so
            that each of these file parts has close to expected size (as
            discussed in Example~\ref{eg:dec} in
            Section~\ref{sec:background}), this results in a rate of 
            \begin{equation*}
                (M/N')(1-M/N')+2(1-M/N')^2 = R(M, N', 2) = 10/9.
            \end{equation*}
            with $R(M,N,K)$ as defined in~\eqref{eq:dec}.  Since all the
            requested files are already partially cached, the set of
            cached files stays the same in the cache update procedure.
            In other words, each user still partially caches
            $\{A,B,C\}$. 

        \item[$t=2$:] The set of popular files changes to $\mc{N}_2 =
            \{B,D\}$. Assume the users request $d_2 = (B,D)$. Here, file
            $B$ is partially cached at the users but file $D$ is not.
            The server sends $B_\emptyset$, $D_\emptyset$, and
            $B_2\oplus D_1$. Since $D$ is not cached at any of the
            users, we have in this case that $D_\emptyset = D$ and $D_1
            = \emptyset$. Hence, the transmission of the server is
            equivalently $B_\emptyset$, $D$ and $B_2$. This results in a
            rate of 
            \begin{equation*}
                (M/N')(1-M/N')+(1-M/N')^2+1 = R(M, N', 1)+1 = 15/9.
            \end{equation*}
            Since $A$ is the least-recently sent file, it is evicted
            from each cache and replaced by a random third of the file
            $D$. The new set of partially cached files is $\{B, C, D\}$. 

        \item[$t=3$:] The set of popular files stays the same $\mc{N}_3
            = \{B,D\}$. Assume the users request $d_3 = (D,B)$, both of
            which are now partially cached at the users.
            The server now sends $B_\emptyset$, $D_\emptyset$, and
            $B_2\oplus D_1$. Unlike the previous time step, $D_1$ is now
            no longer empty, and the resulting rate is
            \begin{equation*}
                R(M, N', 2) = 10/9
            \end{equation*}
            as calculated before. The set of partially cached files
            stays the same, namely $\{B,C,D\}$. 
    \end{description}
\end{example}

It is worth comparing the proposed coded LRS algorithm with the
well-known LRU algorithm described in Example~\ref{eg:lru} in
Section~\ref{sec:problem}. Both of them are online algorithms. However,
there are three key differences.  First, coded LRS uses a coded delivery
procedure whereas the transmissions in LRU are uncoded. Second, coded
LRS caches many ($N'$) partial files whereas LRU caches fewer ($M$)
whole files. Third, coded LRS uses a least-recently sent eviction rule,
taking into account the files requested by all users jointly, compared
to the least-recently used eviction rule, taking into account only the
files requested by every user individually. The impact of these
differences will be explored in more detail later.

\subsection{Theoretical Results}
\label{sec:main_theory}

The main result of this paper is the following theorem.
\begin{theorem}
    \label{thm1}
    The long-term average rate $\bar{R}^\star$ of the optimal online
    caching scheme satisfies
    \begin{equation*}
        \frac{1}{12} R(M, N, K)
        \leq \bar{R}^\star 
        \leq 2 R(M, N, K)+6,
    \end{equation*}
    where
    \begin{equation*}
        R(M, N, K) \defeq K\cdot(1-M/N)\cdot\frac{N}{KM}\bigl(1-(1-M/N)^K\bigr).
    \end{equation*}
\end{theorem}

The proof of Theorem~\ref{thm1} is presented in Section
\ref{sec:proof1}. The upper bound in Theorem~\ref{thm1} results from the
analysis of a simplified version of the proposed coded LRS caching
scheme, showing that this algorithm is approximately optimal. For
the lower bound, we use the rate of the optimal offline scheme,
whose rate is approximately $R(M,N,K)$. 

The theorem thus implies that the rate of the optimal online caching
scheme is approximately the same as the rate of the optimal offline
scheme. Recall that in an offline scheme, the cache memories are given
access to the entire set of popular files each time it changes.
Moreover, these cache updates are performed offline, meaning that the
data transfer needed to update and maintain the caches is not counted
towards the load of the shared link. In contrast, in the online
scenario, caches are updated based on the limited observations they have
through the sequence of demands. Moreover, the cache updates is
performed through the same shared link, and therefore affects the
average rate. Theorem~\ref{thm1} thus indicates that these significant
restrictions on the online problem have only a small effect on the rate
compared to the offline scheme.

\begin{figure}[htbp]
    \centering
    \includegraphics{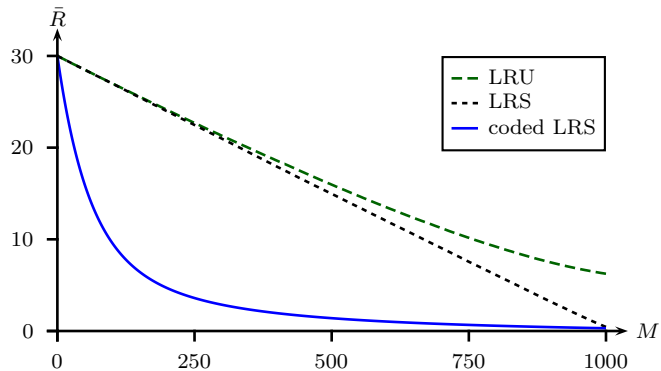}
    \caption{Performance of various caching approaches for a system with
        $N=1000$ popular files, $K=30$ users, and arrival probability $p=0.1$.
        The figure plots the long-term average rate $\bar{R}$ over the
        shared link as a function of cache memory size $M$ for LRU
        (dashed green), uncoded LRS (dashed black), and the proposed
        coded LRS (solid blue).}
    \label{fig:synthetic}
\end{figure}

We now compare the proposed coded LRS scheme to the baseline LRU scheme.
The performances of these two schemes are shown in
Fig.~\ref{fig:synthetic} for a system with $N=1000$ popular files,
$K=30$ users, and arrival probability $p=0.1$. As is visible from the
figure, coded LRS provides significant gains over LRU both for small and
large memory sizes. For example, for $M=250$ (meaning that the cache is
large enough to hold $1/4$ of the popular files), LRU results in a rate
of $22.7$ (meaning that we need to send the equivalent of $22.7$ files
over the shared link on average), whereas coded LRS results in a rate of
$3.6$. Similarly, for $M=1000$, LRU results in a rate of $6.2$, whereas
coded LRS results in a rate of $0.3$. 

As mentioned in Section~\ref{sec:main_algo}, the three main differences
between coded LRS and LRU are coded delivery, partial caching, and LRS
eviction. To get a sense of the impact of these three differences,
Fig.~\ref{fig:synthetic} also depicts the performance of the uncoded LRS
scheme. In this scheme, $M$ whole files are cached and uncoded delivery
is used (as in LRU), however the LRS eviction rule is used (unlike in
LRU).

Comparing (uncoded) LRS to LRU, we see that the two schemes perform
quite similarly for small and moderate values of $M$, say $0 \leq M \leq
750$. For large values of $M$, say $750 < M \leq 1000$, LRS provides a
significant improvement over LRU. This is because when $M$ is close to
the number of popular files $N$, the rate is dominated by the arrival of
new popular files, and LRS eviction allows the caches to learn these new
files with fewer cache misses than LRU.

Comparing uncoded LRS to coded LRS, we see that only when $M$ is very
close to the number of popular files $N$ are the performances of the
schemes similar. For all other values, coded LRS significantly
outperforms uncoded LRS. This implies that, except for large values of
$M$, the main gain of the coded LRS scheme derives from the partial
caching of many files and from the coded delivery.

\subsection{Empirical Results}
\label{sec:main_sim}

To validate the theoretical analysis in Section~\ref{sec:main_theory} as
well as our model for the evolution of popular files, we now evaluate
the performance of the proposed coded LRS and the baseline LRU schemes
for a real-life  time series of demands. This time series is derived
from the dataset made available by Netflix for the Netflix Prize as
follows. Each entry in the Netflix dataset consists  of a user ID, a
movie ID, a time stamp, and a rating that the user gave to the movie at
the specified time. We are not interested in the rating here, but would
like to use the time of rating a movie as a proxy for the time of
viewing that movie. 

This approach is problematic for old movies, which users may rate long
after they have seen them. However, it is reasonable to expect that the
rating time is close to the viewing time for recently released movies.
To ensure that this is the case, we selected all user ratings in the
database from the year 2005 (the last full year for which ratings are
available), and kept only those that are for movies released in either
2004 or 2005. The resulting filtered time series contains about $10^7$
user ratings for $1948$ unique movies.

\begin{figure}[htbp]
    \centering
    \includegraphics{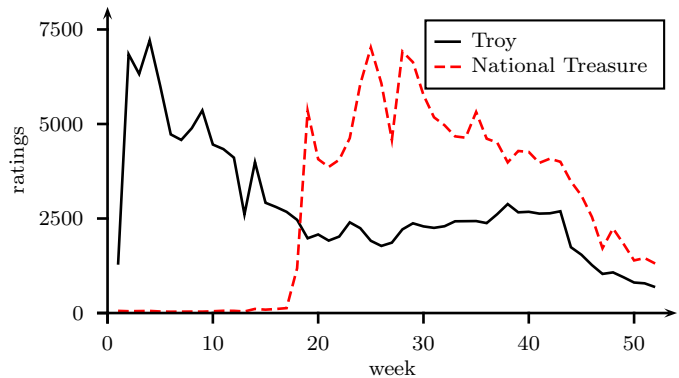}
    \caption{Number of ratings in the Netflix database for two movies
    (``Troy'' and ``National Treasure'') as a function of week in
    2005.}
    \label{fig:troy_treasure}
\end{figure}

To validate this approach, Fig.~\ref{fig:troy_treasure} plots the number
of ratings for the two most-rated movies (``Troy'' and ``National
Treasure'') as a function of time measured in weeks. The movie ``Troy''
was released on DVD on January 4, 2005 (at which time it was likely also
available on Netflix), corresponding to week 1. The movie ``National
Treasure'' was released on DVD on May 3, 2005, corresponding to week 18.
As can be seen from Fig.~\ref{fig:troy_treasure}, the number of ratings
increases strongly on the DVD release week, stays relatively high for a
number of weeks, and then drops. This suggests that the rating time
is indeed a valid proxy for the viewing time when applied to recently
released movies. It also suggests that the model of time-varying popular
files described in Section~\ref{sec:problem} and used for the
theoretical analysis in Section~\ref{sec:main_theory} is a reasonable
model for the viewing behavior of users.

\begin{figure}[htbp]
    \centering
    \includegraphics{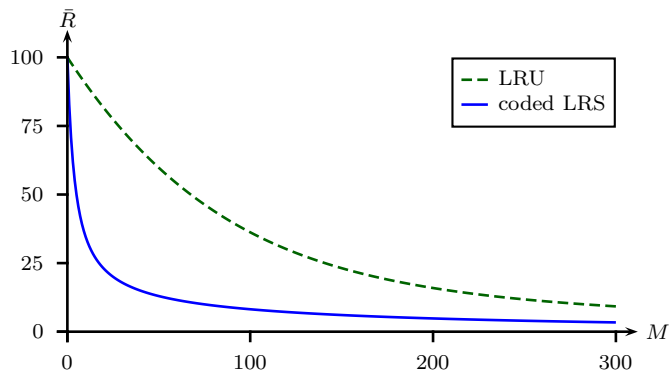}
    \caption{Performance of various caching approaches for the Netflix
        demand time series and a system with $K=100$ caches.
        The figure plots the long-term average rate $\bar{R}$ over the
        shared link as a function of cache memory size $M$ for LRU
        (dashed green) and the proposed coded LRS (solid blue).}
    \label{fig:netflix}
\end{figure}

Fig.~\ref{fig:netflix} compares the performance of the proposed coded
LRS scheme to the baseline LRU scheme for the Netflix demand time series
for a system with $K=100$ caches (each in this case corresponding to
many users that are attached to it). As is visible from the figure,
coded LRS again significantly outperforms LRU. In particular, for a
cache size of $M=100$, LRU achieves a rate of $36.2$ compared to a rate
of $8.2$ for coded LRS.

\section{Proof of Theorem~\ref{thm1}}
\label{sec:proof1}

To prove Theorem~\ref{thm1}, we establish an upper bound
(Section~\ref{sec:proof1_upper}) and a lower bound
(Section~\ref{sec:proof1_lower}) on the optimal long-term average rate
$\bar{R}^\star$.

\subsection{Upper Bound in Theorem~\ref{thm1}}
\label{sec:proof1_upper}

For the upper bound on $\bar{R}^\star$, we analyze a simplified version
of the proposed coded LRS scheme. We refer to this simplified scheme as
\emph{coded random eviction}. In coded random eviction, if $Y_t$ files
are requested by users at time $t$ that are not currently partially
cached, then $ Y_t$ of the cached files are randomly chosen, evicted
from \emph{all} cache memories, and replaced with $MF/N'$ randomly
chosen bits from each of the $Y_t$ newly requested files.  Observe that
this guarantees that the same collection of files is partially cached at
each user.  The remainder of the algorithm is the same as coded LRS as
listed in Algorithm~\ref{alg:1}. In particular the coded random-eviction
algorithm partially caches $N'=\alpha N$ files for a positive constant
$\alpha \geq 1$ at each user, where $N$ is the cardinality of the set of
popular files $\mc{N}_t$. 

Recall that, according to the system model described in Section
\ref{sec:problem}, at each time slot $t$, the users request $K$ randomly
chosen files from the set of popular files $\mc{N}_t$ without
replacement. At any time $t$, not all files in $\mc{N}_t$ may be
partially cached at the users. As in the previous paragraph, we denote
by $Y_t$ the (random) number of uncached files requested by the users at
time $t$.  By definition, $Y_t$ takes value in set $\{0,1,\ldots,K\}$. 

The delivery procedure in Algorithm~\ref{alg:1} transmits these $Y_t$
files uncoded over the shared link.  To send the remaining $K-Y_t$ files
that are partially cached at the users, the delivery procedure of
Algorithm~\ref{alg:1} uses coding. This requires a rate of $R(M, \alpha
N, K-Y_t)$ as described in Section~\ref{sec:background}, and with $R(M,
N, K)$ as defined in Theorem~\ref{thm1} and in~\eqref{eq:dec}.  Thus,
the rate $R_t$ over the shared link at time $t$ is
\begin{align*}
    R_t  
    & = R(M, \alpha N, K-Y_t)+Y_t \\
    & \leq  R(M, \alpha N, K)  +Y_t.
\end{align*}

The long-term average rate $\bar{R}$ of coded random eviction is
therefore upper bounded by
\begin{align}
    \label{eq:ave_r}
    \bar{R} 
    & = \limsup_{T\to\infty} \frac{1}{T}\sum_{t=1}^T\E(R_t), \nonumber\\
    & \leq R(M, \alpha N, K)  
    + \limsup_{T\to\infty} \frac{1}{T}\sum_{t=1}^T\E(Y_t). 
\end{align}
To prove Theorem~\ref{thm1}, we show that the first term
in~\eqref{eq:ave_r} is approximately $R(M, N, K)$ and that the second
term is upper bounded by a constant. This second upper bound is perhaps
surprising, since $Y_t$ itself can take any value up to $K$
and is hence not upper bounded by a constant independent of the problem
parameters. 
  
We start with the analysis of the second term in~\eqref{eq:ave_r}.  Let
the random variable $X_t$ denote the number of files in $\mc{N}_t$ that
are partially stored in the caches at the beginning of time slot $t$.
Note that $X_t$ takes value in $\{0, 1, \dots, N\}$.  Conditioned on
$X_t$, the random variable $Y_t$ has expected value
\begin{equation*}
    \E(Y_t\mid X_t) = K(1-X_t/N).
\end{equation*} 
Therefore, 
\begin{align} 
    \label{eq:x-y}
    \limsup_{T\to\infty} \frac{1}{T}\sum_{t=1}^T\E(Y_t) 
    & = \limsup_{T\to\infty} \frac{1}{T}\sum_{t=1}^T
    \E\bigl( \E(Y_t \mid X_t) \bigl) \nonumber\\
    & = \limsup_{T \to \infty} \frac{1}{T} 
    \sum_{t=1}^T K\bigl(1-\E(X_t)/N\bigr) \nonumber\\
    & = K\biggl(1-\frac{1}{N}\liminf_{T \to \infty} \frac{1}{T}
    \sum_{t=1}^T \E(X_t)\biggr).
\end{align}
In what follows, we investigate the random process $\{X_t\}_{t \in \N}$.

\begin{lemma} 
    \label{thm:ergodic}
    $\{X_t\}_{t \in \N}$ is a Markov process and has a unique stationary
    distribution $\pi = (\pi_0, \pi_1, \dots, \pi_N)$. Moreover,
    \begin{equation*}
        \liminf_{T \to \infty} \frac{1}{T} \sum_{t=1}^T \E(X_t)
        \geq \E(X),
    \end{equation*}
    where $X$ is distributed according to $\pi$. 
\end{lemma}

\begin{IEEEproof}
    Due to the nature of the random-eviction algorithm, and due to  the
    memoryless arrivals and departures to the set of popular files
    $\mc{N}_t$,  $\{X_t\}$ is a Markov process. It is easy to see that
    this Markov process has a single ergodic recurrent class consisting
    of the states $\{\ceil{K/2}-1, \ceil{K/2}, \dots, N\}$ and has
    transient states $\{0, 1, 2, \dots, \ceil{K/2}-2\}$. Therefore,
    $\{X_t\}$ has a unique stationary distribution $\pi$.

    From Fubini's theorem and the properties of $\liminf$, we have
    \begin{align*}
        \liminf_{T \to \infty} \frac{1}{T} \sum_{t=1}^T \E(X_t)
        = \liminf_{T \to \infty} \sum_{x=0}^N x
        \frac{1}{T} \sum_{t=1}^T \Pp(X_t = x) \\
        \geq  \sum_{x=0}^N x \liminf_{T\to\infty} \frac{1}{T} \sum_{t=1}^T \Pp(X_t = x).
    \end{align*}
    Since 
    \begin{equation*}
        \lim_{T\to\infty} \Pp(X_t = x) = \pi_x
    \end{equation*}
    by ergodicity, we obtain from the Ces{\`a}ro-mean theorem that 
    \begin{align*}
        \lim_{T \to \infty}\frac{1}{T} \sum_{t=1}^T \Pp(X_t =x) 
        = \pi_x.
    \end{align*} 
    This implies that
    \begin{equation*}
        \sum_{x=0}^N x \liminf_{T\to\infty} \frac{1}{T} \sum_{t=1}^T \Pp(X_t = x)
        = \sum_{x=0}^N x \pi_x 
        = \E(X), 
    \end{equation*}
    completing the proof. 
\end{IEEEproof}

Applying Lemma~\ref{thm:ergodic} to~\eqref{eq:x-y} yields that
\begin{equation} 
    \label{eq:x-y2}
    \limsup_{T\to\infty} \frac{1}{T}\sum_{t=1}^T\E(Y_t) 
    \leq K\bigl(1-\E(X)/N\bigr).
\end{equation}
To establish the upper bound on Theorem~\ref{thm1}, it thus remains to
lower bound $\E(X)$. This is done in the next lemma.

\begin{lemma}
    \label{thm:quad}
    Let $X$ be as in Lemma~\ref{thm:ergodic}. Then
    \begin{equation*}
        K\bigl(1-\E(X)/N\bigr)  
        \leq \frac{1}{(1-1/N)(1-1/\alpha)}.
    \end{equation*}
\end{lemma}

\begin{IEEEproof}
    To analyze $\E(X)$, we will need a more detailed understanding of
    the random process $\{X_t\}$.  We define two auxiliary processes
    $W_t$, and $U_t$, both for $t \in \N$. $W_t$ is the number of
    randomly evicted files from the caches at the end of time slot $t$
    that are in $\mc{N}_{t}$. In other words, $W_t$ counts the number of
    wrongly evicted files. $U_t$ is the indicator random variable of the
    event that, at the end of time slot $t$, there is a departure from
    the set $\mc{N}_t$ of popular files and that the departing file is
    partially stored in the caches at the end of time slot $t$ (i.e.,
    after the cache update).  
    
    Using these auxiliary processes, we can write the following update
    equation for process $X_t$:
    \begin{equation}
        \label{eq:update}
        X_{t+1} = X_t + Y_t - W_t - U_t.
    \end{equation}
    In words,~\eqref{eq:update} states that the number of correctly
    cached files at time $t+1$ is equal to the number $X_t$ of correctly
    cached files at time $t$, plus the number of newly requested and
    cached files $Y_t$, minus the number of wrongly evicted files $W_t$,
    minus the number of files $U_t$ that were correctly cached at the
    end of time slot $t$ but are no longer popular at time $t+1$.  

    \begin{example}
        \label{eg:update}
        Consider a scenario with $K=3$ users and $N=N'=3$ files.
        Let $\mc{N}_1 = \{C, D, E\}$ and assume at time $t=1$ the files
        $\{A, B, C\}$ are partially cached. Then $X_1 = 1$, since the
        overlap is file $C$. Assume the
        users request $(C, D, E)$ at time $t=1$. Then $Y_1 = 2$, since
        two uncached files $D$ and $E$ are requested. To accommodate the
        two new files, we randomly evict two cached files. Assume those
        files are $B$ and $C$ so that the cached files at the end of
        time slot $t$ are $\{A, D, E\}$. Then $W_1 = 1$, since file
        $C$ in $\mc{N}_1$ is evicted from the caches. Finally assume that file
        $D$ is randomly selected to depart from $\mc{N}_1$ and is replaced
        by the new file $F$, so that $\mc{N}_2 = \{C, E, F\}$. Then $U_t
        = 1$, since the departed file $D$ is cached at the end of time
        slot $t$. Finally, $X_2 = 1$, since file $E$ is both popular
        and cached. This satisfies
        \begin{equation*}
            X_2 = X_1+Y_1-W_1-U_1.
        \end{equation*}
    \end{example}

    To establish a lower bound on $\E(X)$ we use the update
    equation~\eqref{eq:update} instead of directly computing the
    stationary distribution $\pi$, which is not tractable.  Assume that
    the process $\{X_t\}$ is started in steady-state.\footnote{We point
    out that this assumption is merely a proof technique. The actual
    caching system itself may not be started in steady state.} In other
    words, $X_t$ has distribution $\pi$ for every $t\in\N$. Then, taking
    expectations on both sides of \eqref{eq:update}, 
    \begin{equation*}
        \E(X_{t+1}) 
        = \E(X_t) + \E(Y_t - W_t) - \E(U_t).
    \end{equation*}
    Since $\E(X_t) = \E(X) = \E(X_{t+1})$, this simplifies to
    \begin{equation}
        \label{eq:exp0}
        \E(U_t) = \E(Y_t-W_t).
    \end{equation}

    We now calculate the two expectations in~\eqref{eq:exp0}. We start
    with the left expectation. Consider the number of correctly cached
    files $X_{t+1}'$ at the end of time slot $t$ just after the caches
    are updated but before any departures from the set of popular files
    $\mc{N}_t$. Note that
    \begin{equation*}
        X_{t+1}' = X_{t+1}+U_t.
    \end{equation*}
    Conditioned on $X_{t+1}'$,
    \begin{equation*}
        \Pp(U_t = 1 \mid X_{t+1}') = p\frac{X_{t+1}'}{N},
    \end{equation*}
    since a departure happens with probability $p$ and since there are
    $N$ files out of which $X_{t+1}'$ are popular.  Now, $U_t\in\{0,1\}$
    so that
    \begin{align*}
        \E(U_t) 
        & = \E\bigl(\E(U_t \mid X_{t+1}')\bigr) \\
        & = \E\bigl(\Pp(U_t = 1 \mid X_{t+1}')\bigr) \\
        & = p\frac{\E(X_{t+1}')}{N} \\
        & = p\frac{\E(X_{t+1})+\E(U_t)}{N} \\
        & = p\frac{\E(X)+\E(U_t)}{N}.
    \end{align*}
    Solving for $\E(U_t)$ yields
    \begin{equation}
        \label{eq:exp1}
        \E(U_t) = \frac{p}{1-p/N}\frac{\E(X)}{N}.
    \end{equation}

    We then consider the right expectation in~\eqref{eq:exp0}.
    Conditioned on $Y_t$ and $X_t$, the random variable
    $W_t$ has expected value 
    \begin{equation*}
        \E(W_t \mid  X_t, Y_t) = Y_t\frac{X_t}{\alpha N},
    \end{equation*}
    since $Y_t$ files are evicted and $X_t$ of the $\alpha N$ partial
    files in memory are correctly cached. Thus,
    \begin{equation*}
        \E(Y_t-W_t \mid X_t, Y_t) 
         = Y_t\Bigl(1-\frac{X_t}{\alpha N}\Bigr),
    \end{equation*}
    and
    \begin{align*}
        \E(Y_t-W_t \mid X_t) 
         & = \E(Y_t \mid X_t) \Bigl(1-\frac{X_t}{\alpha N}\Bigr) \\
         & = K\Bigl(1-\frac{X_t}{N}\Bigr)\Bigl(1-\frac{X_t}{\alpha N}\Bigr).
    \end{align*}
    Finally, the right expectation in~\eqref{eq:exp0} can be
    evaluated as
    \begin{align}
        \label{eq:exp2}
        \E\bigl( \E(Y_t-W_t \mid X_t) \bigr)
        & = \E\biggl(
        K\Bigl(1-\frac{X_t}{N}\Bigr)\Bigl(1-\frac{X_t}{\alpha N}\Bigr)
        \biggr) \nonumber\\
        & = \E\biggl(
        K\Bigl(1-\frac{X}{N}\Bigr)\Bigl(1-\frac{X}{\alpha N}\Bigr)
        \biggr).
    \end{align}

    For ease of notation, define
    \begin{align*}    
        \bar{x} & \defeq \frac{1}{N}\E(X), \\
        \sigma^2 & \defeq \frac{1}{N^2}\var(X), \\
        \tilde{p} & \defeq \frac{p}{1-p/N}, \\
        \shortintertext{and} \\
        \beta & \defeq 1/\alpha.
    \end{align*}
    Substituting~\eqref{eq:exp1} and~\eqref{eq:exp2} into
    \eqref{eq:exp0} and rearranging yields then
    \begin{align}
        \label{eq:quad2}
        K\beta \bar{x}^2 - (\tilde{p}+K(1+\beta)) \bar{x}
        + K(1+\beta\sigma^2) = 0.
    \end{align}

    This is a quadratic equation in the expected value $\bar{x}$ of
    $X/N$. In Appendix~\ref{sec:appendix_quadratic}, we show that
    the solutions to this quadratic equation can be lower bounded as
    \begin{equation}
        \label{eq:quadratic}
        \bar{x} 
        \geq \frac{\tilde{p}+2K\beta-\tilde{p}(1+\beta)/(1-\beta)}{2K\beta}.
    \end{equation}
    Observe that, crucially, this lower bound does not depend on the
    variance $\sigma^2$ of $X/N$.  Using this lower bound on $\bar{x}$,
    we obtain after some algebra,
    \begin{align*}
        K(1-\E(X)/N)
        & = K(1-\bar{x}) \\
        & \leq \frac{\tilde{p}}{1-\beta} \\
        & = \frac{p}{(1-p/N)(1-1/\alpha)} \\
        & \leq \frac{1}{(1-1/N)(1-1/\alpha)},
    \end{align*}
    concluding the proof of Lemma~\ref{thm:quad}.
\end{IEEEproof}

Applying Lemma~\ref{thm:quad} to~\eqref{eq:x-y2} and substituting
into~\eqref{eq:ave_r} shows that
\begin{equation*}
    \bar{R} 
    \leq R(M, \alpha N, K)  + \frac{1}{(1-1/N)(1-1/\alpha)}.
\end{equation*}
In Appendix~\ref{sec:appendix_ineq}, we show that $R(M, \alpha N, K)$  is
upper bounded as,
\begin{align}
    \label{ineq:2}
    2 R(M,N, K)+\frac{\alpha-1}{1-\alpha/2}.
\end{align}
for $\alpha < 2$. Hence,
\begin{equation*}
    \bar{R} \leq 2 R(M, N, K) + \frac{1}{(1-1/N)(1-1/\alpha)}+\frac{\alpha-1}{1-\alpha/2}.
\end{equation*}
Setting $\alpha = 1.4$, we obtain for $N \geq 7$, 
\begin{equation*}
    \bar{R} \leq 2 R(M, N, K) + 6.
\end{equation*}
On the other hand, for $N \leq 6$, we trivially have
\begin{equation*}
    \bar{R} \leq 6 \leq 2 R(M, N, K) + 6.
\end{equation*}

Since the long-term average of the optimal scheme is less than or equal
to $\bar{R}$, this implies
\begin{equation*}
    \bar{R}^\star \leq \bar{R} \leq 2 R(M, N, K) + 6,
\end{equation*}
concluding the proof of the upper bound in
Theorem~\ref{thm1}.\hfill\IEEEQED

\subsection{Lower Bound in Theorem~\ref{thm1}}
\label{sec:proof1_lower}

We consider a offline scenario in which all cache memories are fully
aware of the set of popular files $\mc{N}_t$. In addition, at the
beginning of each time slot $t$, before users decide on their requests,
the caches are given full access to all the files in $\mc{N}_t$ to
update their stored content at no cost. However, the cache memories are
not aware of future requests. Clearly, the rate of the optimal scheme
for this offline setting is a lower bound on the optimal rate for the
online setting.

This offline problem is in fact equal to the prefetching problem
investigated in~\cite{maddah-ali12a, maddah-ali13, niesen13}, where it
is shown that the instantaneous rate, and therefore also the long-term
average rate, is lower bounded by $1/12 R(M,N,K)$. Thus, 
\begin{equation*}
    \bar{R}^\star \geq \frac{1}{12} R(M,N,K),
\end{equation*}
as needed to be shown.\hfill\IEEEQED

\appendices

\section{Proof of~\eqref{eq:quadratic} }
\label{sec:appendix_quadratic}

Set
\begin{align*}
    a & \defeq \beta K, \\
    b & \defeq -(\tilde{p}+(1+\beta)K\bigr), \\
    c & \defeq (1+\beta\sigma^2)K.
\end{align*}
Then, \eqref{eq:quad2} can be written as $a\bar{x}^2+b\bar{x}+c=0$ with
solutions $\frac{-b \pm \sqrt{b^2 -4ac}}{2a}$.  Since $\bar{x}$ is the
average of a real random sequence and satisfies the above quadratic
equation, this equation has real solutions. In this case, the smaller
solution is with the negative sign. Thus,
\begin{align*}
    \bar{x} 
    & \geq \frac{-b - \sqrt{b^2 -4ac}}{2a} \\
    & = \frac{\tilde{p}+(1+\beta)K-\sqrt{(\tilde{p}+(1+\beta)K)^2 
    - 4\beta (1+\beta\sigma^2)K^2}}{2\beta K} \\
    & \geq \frac{\tilde{p}+(1+\beta)K-\sqrt{(\tilde{p}+(1+\beta)K)^2
    - 4\beta K^2}}{2\beta K},
\end{align*}
where for the last inequality we have used that $\beta \geq 0$.

Now
\begin{align*}
    \bigl(\tilde{p}+(1+\beta)K\bigr)^2 - 4\beta K^2 
    & = \Bigl((1-\beta)K  + \frac{1+\beta}{1-\beta}\tilde{p}\Bigr)^2 
    +\biggl(1 - \Bigl(\frac{1+\beta}{1-\beta}\Bigr)^2\biggr) \tilde{p}^2 \\
    & \leq \Bigl((1-\beta)K  + \frac{1+\beta}{1-\beta}\tilde{p}\Bigr)^2,
\end{align*}
where we have again used $\beta \geq 0$. Thus, 
$\bar{x}$ is lower bounded as 
\begin{equation*}
    \bar{x} 
    \geq \frac{\tilde{p}+2\beta K-\tilde{p}(1+\beta)/(1-\beta)}{2 \beta K}.
\end{equation*}

\section{Proof of~\eqref{ineq:2}}
\label{sec:appendix_ineq}

Assume first that $N/M \geq 1/(2-\alpha)$. Then
\begin{align*}
    R(M, \alpha N, K) 
    & = \bigl(\alpha N/M-1\bigr)\bigl(1-(1-M/(\alpha N))^K\bigr) \\
    & \leq \bigl(\alpha N/M-1\bigr)\bigl(1-(1-M/N)^K\bigr) \\
    & \leq 2 \bigl(N/M-1\bigr)\bigl(1-(1-M/N)^K\bigr),
\end{align*}
where the last inequality holds since
\begin{equation*}
    \alpha N/M - 1 \leq 2 (N/M-1)
\end{equation*}
for $N/M \geq 1/(2-\alpha)$ and $\alpha < 2$. Assume then that $N/M <
1/(2-\alpha)$. Then
\begin{align*}
    R(M, \alpha N, K) 
    & = \bigl(\alpha N/M-1\bigr)\bigl(1-(1-M/(\alpha N))^K\bigr) \\
    & \leq \alpha N/M-1 \\
    & \leq \frac{\alpha-1}{1-\alpha/2}.
\end{align*}
Combining those two inequalities shows that
\begin{equation*}
    R(M,\alpha N, K) \leq 2 R(M,N, K)+\frac{\alpha-1}{1-\alpha/2}.
\end{equation*}

\end{document}